\documentclass[11pt, a4paper]{article}
\pdfoutput=1
\usepackage{jheppub}
\usepackage{graphicx}
\usepackage{epsfig,amsmath,amsthm}
\usepackage{epstopdf}

\newcommand{\be}{\begin{equation}}
\newcommand{\ee}{\end{equation}}
\newcommand{\bea}{\begin{eqnarray}}
\newcommand{\eea}{\end{eqnarray}}
\def\bse{\begin{subequations}}
\def\ese{\end{subequations}}
 \newcommand{\IC}{\mathbb{C}}
 
\def\IZ{\relax\ifmmode\hbox{Z\kern-.4em Z}\else{Z\kern-.4em Z}\fi}
\newcommand{\IS}{\mathbb{S}}

\def\del{{\partial}}

  \def\eps{\epsilon}

 \def\sig{\sigma}
\def\lam{\lambda}

\def\presub{\vspace{.5cm} \noindent}

\def\bi{\begin{itemize}} \def\ei{\end{itemize}}

\def\({\left(} \def\){\right)}
\def\[{\left[} \def\]{\right]}
\def\<{\left<} \def\>{\right>}

\def\om{\omega}

\usepackage[OT2,T1]{fontenc}
\DeclareSymbolFont{cyrletters}{OT2}{wncyr}{m}{n}
\DeclareMathSymbol{\Sha}{\mathalpha}{cyrletters}{"58}
\usepackage{shuffle}

\def\Lie{{\cal L}ie} \def\sig{\sigma}

\title{Color structures and permutations}

\author{Barak Kol and Ruth Shir \\
{\it Racah Institute of Physics, Hebrew University, Jerusalem 91904, Israel} \\
{\tt barak.kol,ruth.shir@mail.huji.ac.il}
}

\abstract{Color structures for tree level scattering amplitudes in gauge theory are studied in order to determine the symmetry properties of the color-ordered sub-amplitudes. We mathematically formulate the space of color structures together with the action of permuting external legs. The character generating functions are presented from the mathematical literature and we determine the decomposition into irreducible representations. Mathematically, free Lie algebras and the Lie operad are central. A study of the implications for sub-amplitudes is initiated and we prove directly that both the Parke-Taylor amplitudes and Cachazo-He-Yuan amplitudes satisfy the Kleiss-Kuijf relations.} 

\begin{document}
\maketitle

\section{Introduction}

 The standard first step in a discussion of perturbative Yang-Mills is the decoupling of color from kinematics, which can be described schematically by \be
 A_{tot} = \sum c_J\, A_J 
  \label{Atot-scheme}
  \ee
where $A_{tot}$ represents the total amplitude for a scattering process with $n$ gluons at some specified loop order, $c_J$ are all the possible color structures depending only on the gluon color, and $A_J$ are  partial sub-amplitudes which depend only on the kinematical data, namely the momenta and the polarizations. 

Two sets of color structures are discussed in the literature. The first is composed of products of the group structure constants $f^{a b c}$ which appear in the Feynman rules, for example \be
 f^{a b r}\, f^{r c s}\, f^{s d e} ~. \ee
 Following the review \cite{DixonRev2011} we shall refer to these as f-expressions. The second involves color ordered traces such as \be
 \mbox{Tr} \( T^a\,  T^b\, T^c\, T^d\, T^e\,\) \ee
 which we shall sometimes call t-expressions, where t stands for trace. The corresponding sub-amplitudes are the more popular color-ordered amplitudes.
 
Our first group of issues is related to the reasons for using two sets of color structures and the relation between them. The seminal work \cite{Witten2003} refers to large $N_c$ planarity in order to describe the color ordered amplitudes, see p. 9. However, it was not specified what happens for low $N_c$, or for groups not belonging to the $SU(N_c)$ family. The space of tree-level color structures for $n$ gluons is known to have dimension $(n-2)!$, as we review later. On the other hand there are $(n-1)!$ possible traces due to the cyclic property. Given that, the excellent review \cite{ElvangHuang2013} refers to the t-expressions as an over complete trace basis (see p. 23). Taken literally this is an oxymoron, while when taken to refer to a dependent spanning set, it appears to imply that the corresponding coefficients, namely the color-ordered amplitudes, are not unique.

The answers to this first set of questions will be seen to be essentially available in the literature, and will be reviewed in section \ref{sec:color}, where the presentation is possibly new.

A second topic is the action of permutations. We may permute (or re-label) the external legs in the expression for a color structure and thereby obtain another color structure. This means that the space of color structures is a representation of $S_n$, the group of permutations. A natural question is to characterize this representation including its character and its decomposition into irreducible representations (irreps). This subject would be the topic of section \ref{sec:permute}.

The paper is organized as follows. In the remainder of the introduction we will survey some of the background. Section \ref{sec:color} will present the relation between the f-expressions and the t-expressions. In section \ref{sec:permute} we formulate the mathematical problem of characterizing the $S_n$ representations and we present results, including a list of irreps in appendix \ref{sec:irrep}. The ultimate objective in studying the color structures is to discern the symmetry properties of the colorless amplitudes. In section \ref{sec:implications} we take a first step in this direction. Finally, we conclude in section \ref{sec:conclude}. 

\presub {\bf Background}. For some time it is known that the standard perturbation theory based on Feynman diagrams is not the optimal theory for (non-Abelian) gauge theory, gravity and possibly additional theories, especially for a large number of particles and/or small violation of helicity. Here we wish to present a brief and modest survey of the subject's evolution.

In the sixties \cite{DeWitt1967} observed unexpected cancellations in the 4 particle amplitudes of gravity and Yang Mills. Helicity spinors, whose evolution is described in the review \cite{ManganoParke1990}\footnote{On p. 7 of the arXiv version.}, were used in the eighties to obtain an inspiring simple expression for Maximal Helicity Violating (MHV) amplitudes,  discovered in \cite{ParkeTaylor1986} and proved in \cite{BerendsGiele1987b}. The decomposition  of color structures into irreps was suggested in \cite{Zeppenfeld88}.

Progress in the nineties included unitarity (or on-shell) cuts, and ideas from supersymmetry and string theory, see the review \cite{BernDixonKosower1996} and references therein. 

In 2003 \cite{Witten2003} reignited the interest in this field by suggesting an approach involving a string theory in Twistor space \cite{PenroseTwistors1967}.  One of the outcomes was a discovery of the recursion relations of \cite{BCFW2005}. An intriguing color-kinematics duality introduced in \cite{BCJ2008} reduced the number of independent sub-amplitudes to $(n-3)!$ (the coefficients in the relations are kinematics dependent) and this duality was applied to gravitational amplitudes. 

A search for the underlying geometry, and especially that underlying dual superconformal invariance of ${\cal N}=4$, led to the Grassmannian \cite{ArkaniHamedCachazoCheung2009}, the space of $k$ planes in $n$ dimensional space, where $k$ is related to the number of negative helicity gluons and $n$ is the total number of external gluons. This was successively refined into the positive Grassmannian \cite{PositiveGrassmannian2012}, a subset of the Grassmannian which generalizes the simplex, and most recently into the Amplituhedron \cite{Amplituhedron2013}, which is a larger object which can be triangulated by positive Grassmannians. 

A simple formula for gravitational tree-level MHV amplitudes was presented in \cite{Hodges2012}. The Gross-Mende scattering equations were found to play a central role and allowed \cite{CachazoHeYuan2013} to obtain a remarkable closed formula for all tree level amplitudes in both (pure) gauge theory and gravity, without reference to helicity violation or helicity spinors.

Excellent and relatively recent reviews include \cite{DixonRev2011,ElvangHuang2013,DixonRev2013}. See \cite{KeppelerSjodajl12,EdisonNaculich12} for relatively recent papers on irrep decomposition of color factors with explicit results for $n \le 6$ and \cite{Kanning:2014maa,Broedel:2014pia} which discuss tree-level amplitudes for ${\cal N}=4$ via integrability.

Altogether several ingenious ideas were introduced so far to facilitate computations in perturbative gauge theory. It is quite surprising that so much novelty was found in a topic as old and as heavily studied as perturbative field theory. This multitude of ideas also highlights the current lack of a single unifying framework which one might expect to underlie them.

\section{Color trees and traces}
\label{sec:color}

In this section we define the color structures using two decompositions, one based on the structure constants $f^{abc}$ and the other based on traces. We describe some basic properties including the dimension, the relation between the two decompositions, the derivation of the Kleiss-Kuijf relations \cite{KleissKuijf1988} and the shuffle and split operations, see also \cite{ReuschleWeinzierl13}. Along the way we shall answer the first group of issues mentioned in the introduction. The material in this section is not new, except possibly for presentation, and it is included here in order to set the stage for the next sections. The structure constant decomposition was developed in \cite{Cvitanovic-etal1981,DelDucaDixon-etal1999a,DelDucaDixon-etal1999b}, while the trace based decomposition was developed in \cite{BerendsGiele1987a,ManganoParkeXu1987,Mangano1988} using an analogy with the Chan-Paton factors of string theory \cite{ChanPaton1969}. 

\subsection{f-based color trees}

We consider pure Yang-Mills theories with gauge group $G$, and processes with $n$ external gluons, but limiting ourselves in this paper to the tree-level, which is already interesting. In order to determine the (total) amplitude the following data is needed: for each external gluon we must specify a color $a$  in the adjoint of $G$, its energy-momentum $k^\mu$, and its polarization $\eps^\mu$, altogether listed by \be
\left\{ a,\, k^\mu,\, \eps^\mu \right\}_i, \qquad i=1,\dots,n
\label{def:data}
\ee
where for incoming particles $k$ is reversed and $\eps$ is complex conjugated.

Since the color and kinematic data belong to different spaces we can decompose the total amplitude as follows \be
A_{tot} = \sum_J c_J\, A_J
\label{Atot}
\ee
where $J$ runs over a list of indices to be specified later, the color factors $c_J=c_J(a_i)$ depend only on the color data, and the sub-amplitudes, or partial amplitudes, $A_J=A_J(k_i,\eps_i)$ depend only on the kinematical data. This is the statement of \emph{color -- kinematics decomposition}. The only coupling between color and kinematics is through the particle number labels $i=1,\dots,n$ which they both share, and hence the $J$ indices depend on them. 

Our purpose is to study the space of color structures $c_J$ in order to obtain the number of partial amplitudes $A_J$ and their symmetry properties.

\presub {\bf Color Feynman rules}. The Yang-Mills Feynman rules for the propagator and the cubic vertex factorize to a product of color and kinematics $c \cdot A$. For our purposes the color dependence suffices and the corresponding Feynman rules are \be
  \parbox{20mm}{\includegraphics[scale=0.5]{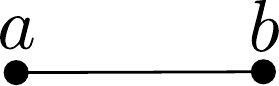}} 
  = \delta_{a b} \qquad
 \parbox{20mm}{\includegraphics[scale=0.5]{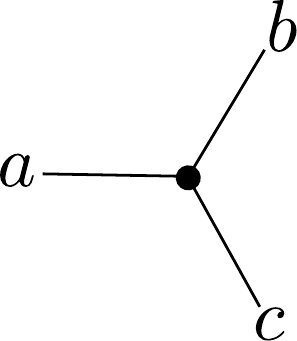}}
= f^{a b c}
\label{def:color-Feynman-rules}
\ee
 where $a,b,c$ are color indices and $f^{a b c}$ are the group structure constants.\footnote{
 They are defined by $\left[ T^a,\, T^b \right] = i\, f^{a b c}\, T^c$, where $T^a$ are the (Hermitian) generators (see for example the textbook \cite{PeskinSchroeder} eq. (15.44) on p. 490). Their normalization is immaterial in this paper. The full (tree level) Feynman rules can be found for example in  \cite{PeskinSchroeder} eq. (16.5) and fig. 16.1 on p. 506-8.}
These rules can be redefined by a multiplicative factor (which can be shifted from $c$ to $A$). They were presented in \cite{Cvitanovic-etal1981}.

It should be stressed that the cubic vertex is \emph{oriented}, namely it specifies a cyclic order of its legs. This is conveniently done through embedding the diagram in the plane and inheriting the orientation from the plane's clockwise orientation.\footnote{The clockwise convention would be convenient to avoid minus signs later in (\ref{def:cf}).} The origin of the orientation is the fact that while the full Feynman rule $c \cdot A$ is invariant under permutations of its legs (or equivalently under a relabeling of vertices), each of its factors is in fact anti-symmetric under such a transformation.

The Yang-Mills quartic vertex is given by a sum over 3 factorized terms \bea
\parbox{20mm}{\includegraphics[scale=0.5]{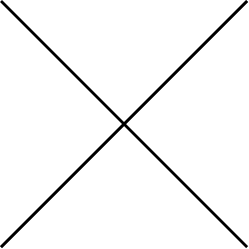}} = \parbox{20mm}{\includegraphics[scale=0.5]{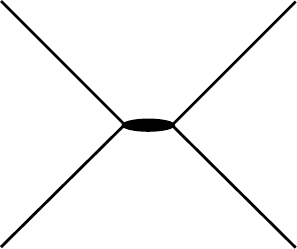}} + \parbox{20mm}{\includegraphics[scale=0.5]{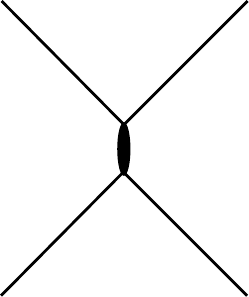}} + \parbox{20mm}{\includegraphics[scale=0.5]{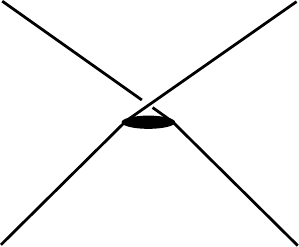}} \\
 \parbox{20mm}{\includegraphics[scale=0.5]{quart1.pdf}} = \parbox{20mm}{\includegraphics[scale=0.5]{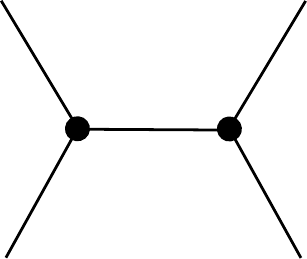}} 
\label{quartic-vertex}
\eea 
where the color factor of each term is the same as that of the analogous diagram with two cubic vertices. Hence, for the purpose of computing all possible color factors quartic vertices need not be considered. With these conventions each Feynman rule factorizes to a product of color and kinematics $c \cdot A$, and hence the full diagram also enjoys the same factorization, see also \cite{BjerrumBohr:2012mg}.

\presub {\bf The diagrammatic representation of the Jacobi identity}. While studying color factors we wish to identify expressions only if they are equal for any Lie algebra, namely only by using the Jacobi identity, which is of course $\[ \[ T^a,\, T^b\], T^c\] + \[ \[ T^c,\, T^a\], T^b\] +\[ \[ T^b,\, T^c\], T^a\] =0$ or equivalently $f^{a b x}\, f^{x c d} + f^{c a x}\, f^{x b d}+ f^{b c x}\, f^{x a d}=0$. It can be described diagrammatically by \be
\parbox{25mm}{\includegraphics[scale=1]{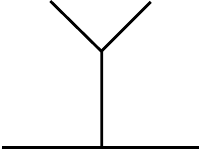}} = \parbox{25mm}{\includegraphics[scale=1]{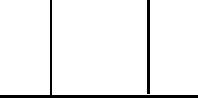}} - \parbox{25mm}{\includegraphics[scale=1]{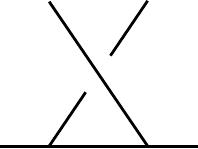}} 
\label{Jacobi}
\ee
as presented in \cite{Cvitanovic76} (also known as an IHX relation \cite{BarNathan95}, see p.42 in the book \cite{CvitanovicGroupTheory08}). 

\presub {\bf Color structures}. The space of tree-level color structures $TCS_n$ may now be defined as the vector space generated by all diagrams with $n$ external legs and an oriented cubic vertex, which are connected and without loops, namely having a tree topology, where diagrams which differ by the diagrammatic form of the Jacobi identity (\ref{Jacobi}) are to be identified. $TCS_n$ is the main object of study of this paper.

\begin{figure}[t!]
\centering \noindent
\includegraphics[width=15cm]{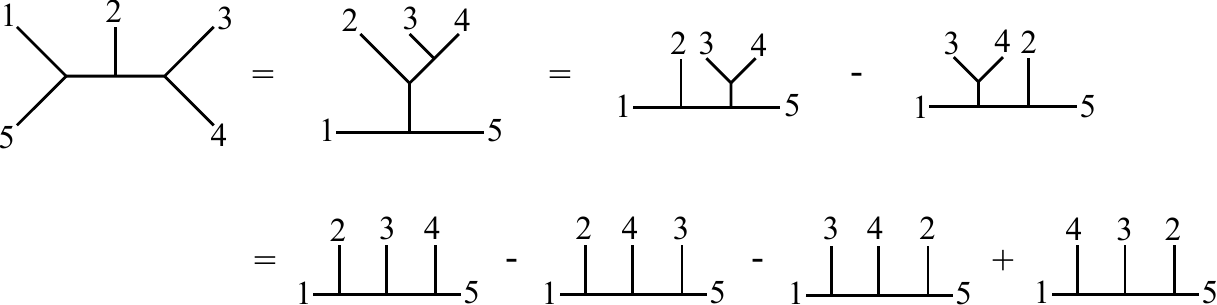}
\caption[]{An example of Jacobi planting, whereby any color diagram can be converted to a combination of flat or multi-peripheral diagrams through a repeated use of the Jacobi identity (\ref{Jacobi}).}
 \label{fig:Jacobi-planting}
\end{figure}

\presub {\bf Dimension of $TCS_n$ through ``Jacobi planting''}. Let us see that \be
 \mbox{dim}(TCS_n) = (n-2)! 
 \label{dim}
 \ee
There are $(2n-5)!! \equiv 1 \cdot 3 \dots (2n-5)$ tree diagrams with cubic vertices.\footnote{
 This can be seen by noting that such an $n$ tree is gotten from an $(n-1)$ tree exactly by attaching an external leg to one of the $2(n-1)-3$ edges.}
 One starts by freely choosing two of the external legs, for instance $1,\, n$.  In any given tree there is a well-defined path connecting $1$ and $n$. One draws that path as a baseline for the diagram. Then by repeated use of the Jacobi identity (\ref{Jacobi}) each subtree emanating from this baseline can be converted to (a linear combination of) separate branches, see for example fig. \ref{fig:Jacobi-planting}. In this way any color factor can be represented by a combination of flat diagrams, or ``multi-peripheral diagrams'' \cite{DelDucaDixon-etal1999b}. The number of flat diagrams equals the number of ways of ordering the remaining external legs $2,\dots,n-1$, which is $(n-2)!$.  The essence of this argument was presented in \cite{Cvitanovic-etal1981} for diagrams with a quark line and $n-2$ gluons and was generalized in \cite{DelDucaDixon-etal1999b} to $n$ gluons. To complete the proof of (\ref{dim}) one needs to show that the flat diagrams do not only span $TCS_n$ but are also linearly independent, as is known to be the case. 
 
\presub {\bf f-based decomposition}. We denote the color factors of flat diagrams by \be
c^f(1 \sigma n) := 
\parbox{50mm}{\raisebox{1em}{\includegraphics[scale=0.6]{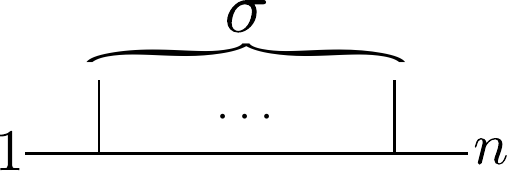}}} \hspace{-2cm} \equiv f^{1 \sig_1 x_1}\, f^{x_1 \sig_2 x_2} \dots f^{x_{n-4} \sig_{n-3}\, x_{n-3}} f^{x_{n-3} \sig_{n-2} n} 
\label{def:cf}
\ee
where $\sigma \in S_{n-2}$, the permutation group on $n-2$ elements, acting here on $2,\dots,n-1$, and the $f$ superscript stands for the structure constants $f^{a b c}$. Now we can introduce a specific color -- kinematics decomposition (\ref{Atot}), known as the f-based decomposition \cite{DelDucaDixon-etal1999a,DelDucaDixon-etal1999b}, and the associated partial amplitudes $A^f(1 \sigma n)$ \be
A_{tot} = \sum_{\sigma \in S_{n-2}} c^f(1 \sigma n)\, A^f(1 \sigma n) ~.
\label{def:f-decomp}
\ee
This decomposition was first discovered in the context of multi-Regge kinematics, namely a monotonic hierarchy of rapidities between all particles, where the dominant sub-amplitude is the $A^f$ whose ordering is determined by rapidity \cite{DelDucaDixon-etal1999a,DelDucaDixon-etal1999b}.

\subsection{Traces}

Color structures have another representation, one which is trace based. To define it each appearance of the structure constants is replaced according to \be
 i\, f^{a b c} \to {\rm Tr}\( T^a\, T^b\, T^c - T^c\, T^b\, T^a\) 
 \label{f-to-tr}
 \ee 
  where the trace is taken in the fundamental representation. In fact, the fundamental representation is not special, and could be replaced by any other representation at the cost of a multiplicative constant. The essential structure is that of the Killing metric on the Lie algebra, which is unique up to a constant. This replacement makes use of products of generators, which are not part of the structure of a Lie algebra, but appear in the enveloping associative algebra.\footnote{
We recall that each Lie algebra ${\cal L}$ is naturally associated with an associative (matrix-like) algebra generated by the same generators and satisfying only the constraints \be
T^a \cdot T^b - T^b \cdot T^a  - \[ T^a, T^b \] = 0
\label{def:enveloping}
\ee
for any two generators $T^a, T^b$, where $[T^a, T^b]$ is their Lie bracket. 
 This algebra is known as the enveloping algebra ${\cal A}_0$,
 and it is the same familiar construction used in quantum mechanics to turn the Lie algebra structure of Poisson brackets into commutators, see for example the classic textbook \cite{DiracQM}.
 Formally the enveloping algebra is given by the free associative tensor algebra generated by the Lie algebra ${\cal L}$,  divided by the ideal generated by the relations (\ref{def:enveloping}). ${\cal A}_0$  is universal in the sense that any Lie homomorphism from ${\cal L}$ to an algebra ${\cal A}$ can be factorized to proceed through ${\cal A}_0$, namely ${\cal L} \to {\cal A}_0 \to {\cal A}$.
}

Next one performs the summations over internal legs. In several excellent reviews \cite{DixonRev2011,ElvangHuang2013,DixonRev2013} this is done for $G=SU(N_c)$ by the completeness identity \be
 \( T^a \)_i^j\, \( T^a \)_k^l = \delta_i^l \, \delta_k^j\, - \frac{1}{N_c}\, \delta_i^j\, \delta_k^l ~.
\label{SUn-completeness}
\ee
In theories with adjoint matter only the the $1/N_c$ term can be omitted.

In fact, for tree diagrams, there is an alternative method which is valid for arbitrary $G$ \cite{ManganoParkeXu1987} and moreover it does not require to introduce an auxiliary representation (such as the fundamental in (\ref{SUn-completeness}) ) for theories with adjoint fields only. One starts by applying (\ref{f-to-tr}) to an arbitrarily chosen vertex and then one applies to adjacent vertices the alternative identity \be
\sum_a  {\rm Tr} \( P\, T^a \) f^{a b c} \to   {\rm Tr} \( P\, \( T^b\, T^c - T^c\, T^b \) \) 
\label{t-concat}
\ee
where $P$ is any product of generators (or a polynomial in generators). This step is repeated until all occurrences of $f^{a b c}$ are converted. In this way any tree level f-expression is transformed into a combination of single trace expressions, which will be called a t-expression (t stands for trace). The same goal can be achieved by using (\ref{f-to-tr}) on all  occurrences of $f^{a b c}$ and then  applying (\ref{SUn-completeness}) to concatenate products of traces into a single trace. This method generalizes to loops, but is special for $G=SU(N_c)$.



Substituting the f color expressions by t-expressions in the f-based  color -- kinematics decomposition (\ref{def:f-decomp}) defines trace-based partial amplitudes $A^t(\sigma_n) ~ \sigma_n \in S_n$ and the corresponding decomposition \cite{BerendsGiele1987a,ManganoParkeXu1987,Mangano1988} \be
 A_{tot} = \sum_{\sigma_{n-1} \in S_{n-1}} {\rm Tr}(\sigma_{n-1} n )\, A^t(\sigma_{n-1} n) ~
\label{def:tr-decomp}
\ee
where the cyclic property of the trace was used to fix the last leg to be $n$. $A^t(\sigma_n)$ are known as color-ordered amplitudes.

{\bf Further motivation}. The trace-based color factors need not be supplemented by the Jacobi identity (since $f^{a b c}$, or equivalently the Lie bracket, is replaced by a commutator).  Whereas \cite{BerendsGiele1987a} arrived at the trace-based color structures by regrouping Feynman diagrams, \cite{ManganoParkeXu1987} were inspired to these quantities by string theories. Indeed, a string diagram for the scattering of $n$ open strings  is proportional to the Chan-Paton color factor \cite{ChanPaton1969} \be
 \parbox{30mm}{\includegraphics[scale=0.8]{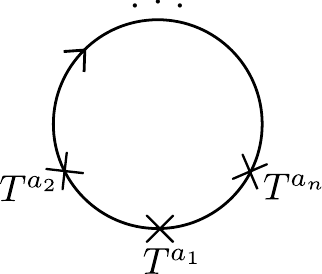}}
 \propto {\rm Tr} \(T^{a_1} \cdot \dots \cdot T^{a_n} \)
\ee
and each segment of the boundary is labelled by one of the $N_c$ space-time filling D-branes. The moduli space of each of these string diagrams includes in it as limits all the field theory Feynman diagrams for the corresponding color ordered amplitude. 

\subsection{Relating the f and trace decompositions}

Let us return to examine in more detail the transformation from f-based to t-based color structures, and in the process we shall identify the corresponding relation between the sub-amplitudes, as well as a certain identity between the color-ordered sub-amplitudes.

We start with mathematical preliminaries, defining the split and shuffle of words. The shuffle is a standard term, also known as the ordered permutation, and the split is our term, sometimes known as the co-shuffle, yet we are unaware of a standard term. For reference see for example the textbook \cite{Reutenauer}. Consider an alphabet, which in our case would be $\{a_i\}$, the color generators of the external legs, and consider the words $w$ that can be made out of it, and their linear combinations, namely the non-commutative polynomials $P$ (which are called polynomials in short, when no confusion can arise). Given a word $w$ its split is the sum of all possible ways to split it into two words, each one preserving the original ordering of the letters. More precisely, the split of a word $\delta(w)$ may be defined recursively by \be
\delta(a\, w) = \[ (a,\emptyset) + (\emptyset,a) \] \delta(w)
\label{def:split}
\ee
where $a$ is any letter, $w$ is any word, $(u,v) \equiv u \otimes v$ and $\emptyset $ denotes the empty word, which serves as a unit for the word algebra. For example \be
 \delta(a b)=(a b,\emptyset)+(a,b)+(b,a)+ (\emptyset,ab) ~.
 \label{ex:split}
 \ee
The split operation naturally extends to polynomials by linearity. Differently put, split is the unique operation respecting polynomial addition and multiplication (concatenation) and having $\delta(a) = (a,\emptyset) + (\emptyset,a)$. 

Not all polynomials $P$ can be written in terms of Lie brackets alone (and no products). Those which do are called Lie words and they can be neatly characterized by the split operation which we just defined, namely a polynomial $P$ is a Lie word if and only if $\delta(P)=(P,0)-(0,P)$.

The shuffle of two words $u$ and $v$, denoted by  
 $u \shuffle v$,\footnote{
 Inspired by the Cyrillic letter Sha $\Sha$, which it turn originates in the Phoenician Shin, and is similar to the modern Hebrew Shin.}
  is the sum of all the ways to interleave $u,v$ into a single word while respecting the ordering within $u,v$. For example \be
a b\, \shuffle \, x y = a b x y + a x b y + a x y b + x a b y + x a y b + x y a b
\label{ex:shuffle}
\ee
Shuffle also extends to all polynomials by linearity.

The split and shuffle are adjoint operators,\footnote{
With respect to the natural inner product on non commutative polynomials on the alphabet. It is  defined such that $P \cdot w$ is the coefficient of the word $w$ in the polynomial $P$ and hence $P=\sum_w (P \cdot w) w$ and it is extended by linearity to any two polynomials $P,Q$ so that $P \cdot Q = \sum_w (P \cdot w) (Q \cdot w)$, see for example \cite{Reutenauer} p.15,17.}
 namely \be
\delta(w) \cdot (u,v) = w \cdot (u \, \shuffle \, v) ~.
\label{split-shuffle-duality}
\ee
In particular $w$ is in the shuffle of $u$ and $v$ if and only if $(u,v)$ is in the split of $w$. 

Now we turn to relating the f-based and t-based decompositions. Our argument will use a graphical representation closely related to \cite{DixonRev2013}. The transformation (\ref{f-to-tr}) is represented by \be
\parbox{70mm}{\includegraphics[scale=0.5]{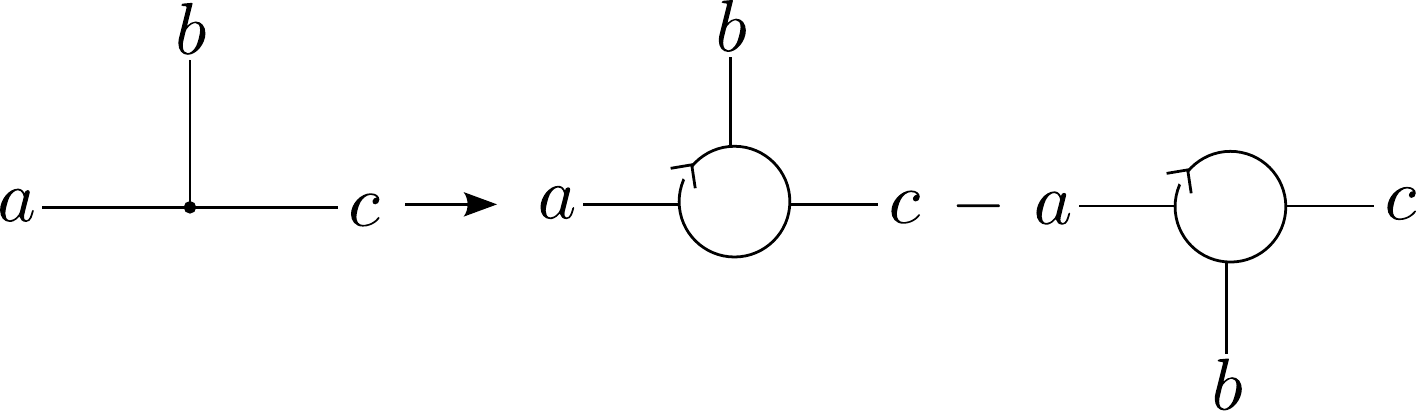}}
\label{graph-f-to-tr}
\ee
 (\ref{SUn-completeness}) is represented by \be 
\parbox{50mm}{\includegraphics[scale=0.8]{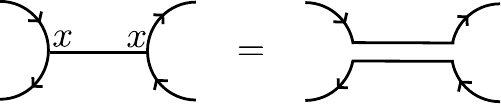}} 
\label{graph-completeness}
\ee
The arrows denote the fundamental representation as in (\ref{f-to-tr}).  We can also represent  (\ref{t-concat}) graphically by \be
 \parbox{50mm}{\includegraphics[scale=0.5]{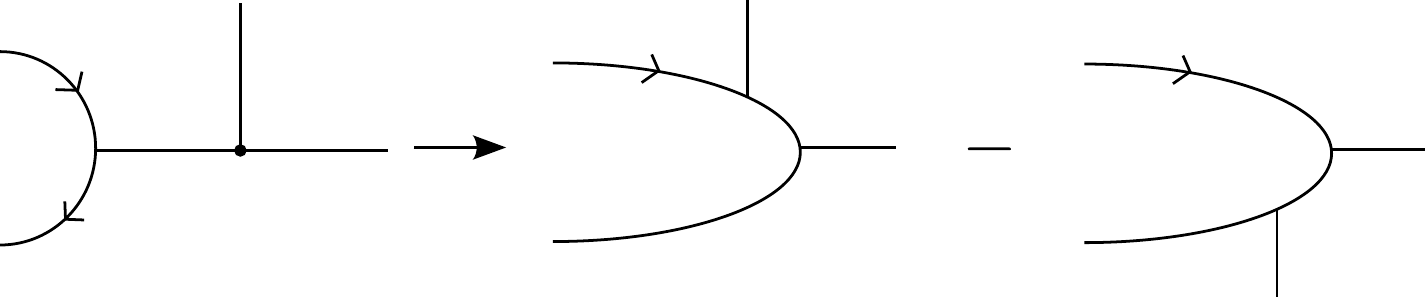}} 
\label{graph-t-concat}
\ee

Before addressing a general color structure it is instructive to consider a color structure $ \sim f^2$. It can be transformed as follows \bea
f^{a b x}\, f^{x c d} &=& \parbox{30mm}{\raisebox{2em}{\includegraphics[scale=0.5]{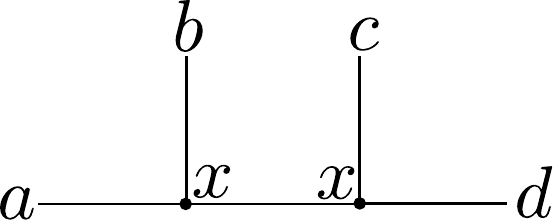}}}
  =  \Big(\parbox{40mm}{\includegraphics[scale=0.45]{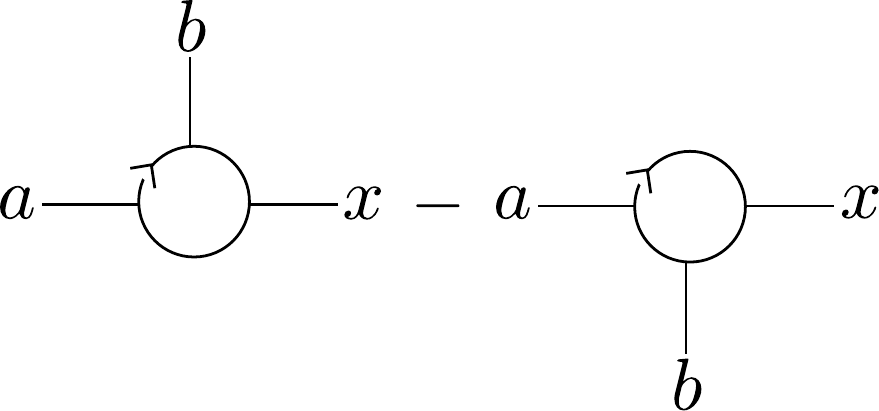}}\Big) \Big(\parbox{40mm}{\includegraphics[scale=0.45]{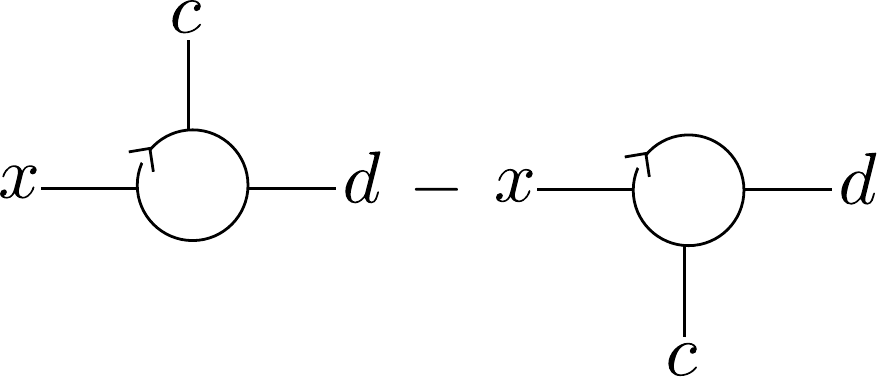}}\Big)
 \nonumber \\
  &=& \parbox{30mm}{\raisebox{2em}{\includegraphics[scale=0.5]{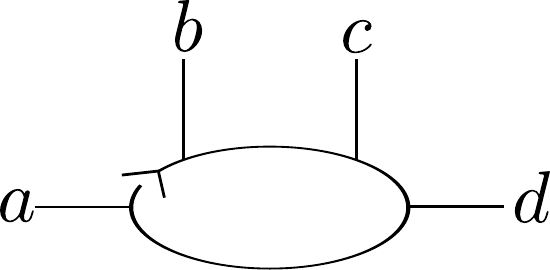}}}
-
 \parbox{30mm}{\includegraphics[scale=0.5]{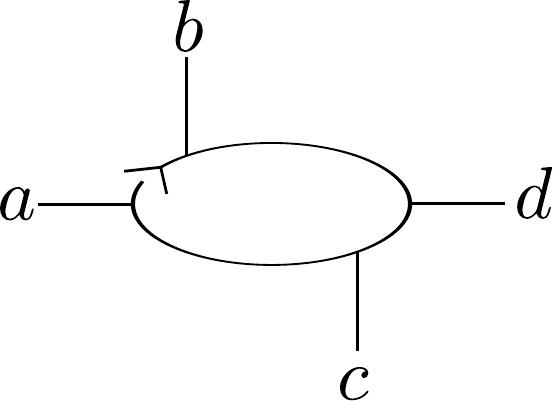}}
  - \parbox{30mm}{\includegraphics[scale=0.5]{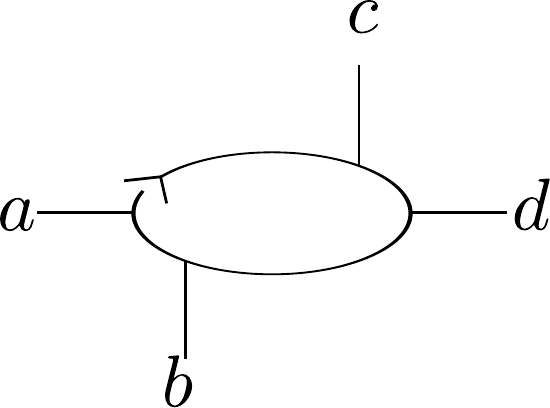}}
 +
 \parbox{30mm}{\raisebox{-5em}{\includegraphics[scale=0.5]{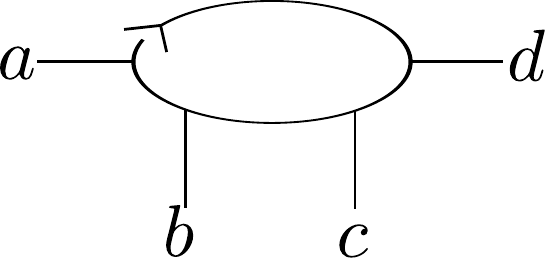}}}\nonumber
 \eea
 
For a general flat color diagram each one of its $n-2$ vertices transforms into a pair of terms according to (\ref{f-to-tr}), such that each of the vertical lines labelled $2,3,\dots,n-1$ can point either upward or downward, resulting with a total of $2^{n-2}$ terms as follows \bea
 c^f(1 2 \dots n) & \equiv & \parbox{30mm}{\raisebox{1em}{\includegraphics[scale=0.55]{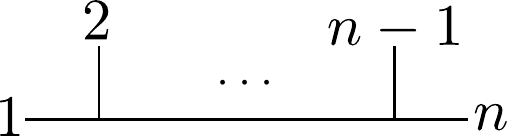}}}
\equiv f^{1 2 x_1}\, f^{x_1 3 x_2} \dots f^{x_{n-3} n-1\, n} \\
 &=& \sum_{(\alpha,\beta) \in \delta(\sigma)} (-)^{|\beta|} \mbox{Tr} \( 1 \alpha n \beta^T \)
 \label{cf-to-ct}
 \eea
where the sum is over the split of $\sigma=2 \dots n-1$, $\alpha$ is the subset of $\sigma$ which points upward, while $\beta$ is the subset which points downward and hence is read in reverse order and is accompanied by a minus sign, and accordingly $|\beta|$ is the size of the set $\beta$, and $\beta^T$ is its transpose, or reflection.

Now the relation between the two forms of color structures can be used to derive the corresponding relations between the sub amplitudes \bea
 A_{tot} &=& \sum_\sigma c^f(1 \sigma n)\, A^f(1 \sigma n)  = \\
 	&=& \sum_\sigma \[ \sum_{(\alpha,\beta) \in \delta(\sigma)} (-)^{|\beta|} \mbox{Tr} \( 1 \alpha n \beta^T \) \] A^f(1 \sigma n) = \\
	&=&   \sum_{(\alpha,\beta)} \[  \sum_{\sigma: (\alpha,\beta) \in \delta(\sigma)} (-)^{|\beta|}  A^f(1 \sigma n) \]  \mbox{Tr} \( 1 \alpha n \beta^T \) = \\
	&=&  \sum_{(\alpha,\beta)} \[  \sum_{\sigma \in \alpha \shuffle \beta } (-)^{|\beta|}  A^f(1 \sigma n) \]  \mbox{Tr} \( 1 \alpha n \beta^T \)
\eea 
 where we used in the first line  (\ref{def:f-decomp}), (\ref{cf-to-ct}) in passing to the second, a change of summation in passing to the third and finally (\ref{split-shuffle-duality}) in the fourth. Comparing with (\ref{def:tr-decomp}) we find \be
  A^t (1 \alpha n \beta^T) = \sum_{\sigma \in \alpha \shuffle \beta } (-)^{|\beta|}  A^f(1 \sigma n) ~.
  \label{Af-to-At}
 \ee
In particular for $\beta=\emptyset$ \be
 A^t(1 \sigma n) = A^f (1 \sigma n)
\label{ft-equiv}
\ee
 and hence the superscripts $f$ and $t$ can and will be omitted henceforth, and $A^f$ turns out to have a cyclic symmetry which is not apparent from its definition. For general $\beta$ and omitting the $t,f$ superscripts (\ref{Af-to-At}) concludes the derivation of the celebrated Kleiss-Kuijf (KK) relations \cite{KleissKuijf1988}.  The results (\ref{Af-to-At},\ref{ft-equiv}) were obtained originally in \cite{DelDucaDixon-etal1999b}.

See \cite{BjerrumBohrDamgaardVanhove09} for a generalized identity in string theory which reduces to the Kleiss-Kuijf identities in the low energy limit.
 
\subsection{Section conclusions} 
\label{subsec:interim-summ}

Let us summarize the discussion in this section. The space of tree level color structures with $n$ external legs, $TCS_n$, has dimension $(n-2)!$. Accordingly $(n-2)!$ partial amplitudes suffice to specify any total amplitude. However, there is no natural, or canonical, basis for this space. Rather, there is a natural set of bases  each one labelled by a choice of initial and final external legs, so there are $n(n-1)$ natural bases, and the total set of color structures is given by $c^f(i \sig j), ~ \sig \in S_{n-2}$. Hence it initially appears that the total number of the associated partial amplitudes would be $n(n-1) \cdot (n-2)!=n!$. However, one can pass to the trace expressions,  which are manifestly cyclic and hence are $(n-1)!$ in number. The trace-based decomposition turns out to define sub-amplitudes which are equivalent to the f-based sub-amplitudes (\ref{ft-equiv}). Hence altogether \emph{there are $(n-1)!$ sub-amplitudes, and $n(n-1)$ ways to choose a basis of $(n-2)!$ sub-amplitudes out of them}.

Let us return to the first group of issues mentioned in the introduction. It is now clear that at least at tree level the existence of color ordered amplitudes is unrelated to the large $N_c$ limit or the gauge groups $SU(N_c)$, and in fact the discussion holds for a general Lie group. It is also clear now that the partial amplitudes are not components in one basis, nor for some spanning set, but rather these are the components for a natural set of bases.
 
{\bf Reflections}. We end this section with a comment on the action of reflections on amplitudes. From the definition of the f-based color factors (\ref{def:cf}) $c^f \sim (f)^{n-2}$ and it is evident that under reflection, $(1\dots n) \to (n \dots 1)$, they transform as $c^f \to (-)^{n-2} c^f = (-)^n\, c^f$ and hence by (\ref{def:f-decomp},\ref{ft-equiv})  
so do the amplitudes, namely \be
 A(1\dots n) \to (-)^n\, A (n \dots 1) ~.
\label{reflect}
\ee
This means that ignoring signs there are only $(n-1)!/2$ different partial amplitudes. However, the dimension of $TCS_n$ is $(n-2)!$ and not half of it, since reflection relates two amplitudes from different bases as it exchanges end-points of the flat diagrams (\ref{def:cf}).

\section{Action by permutations}
\label{sec:permute}

In this section we study the symmetries of the tree level color structures $TCS_n$ with respect to permutations of the external legs, having in mind 
their application to the determination of partial amplitudes.

\presub {\bf Definition of group action}. The natural operation of $S_n$, the group of permutations of $n$ elements, on $TCS_n$ can be defined as follows. Given a color structure, in either an f or trace representation, we can obtain a new one by permuting the labels $1 \dots n$. For example, for $n=4$ the permutation $(1234) \to (1324)$ transforms $f^{1 2 x}\, f^{x 3 4}$ into $f^{1 3 x}\, f^{x 2 4}$. In fact, in this way any permutation induces a linear transformation on $TCS_n$. Alternatively and equivalently, one can take the passive point of view according to which the permutation is a re-labeling of the external legs.

$TCS_n$ is  $(n-2)!$ dimensional, and as such it reminds us of certain basic representations of $S_n$. The regular representation acts on the space of all orderings of $1,\dots,n$ (or permutations), and is therefore $n!$ dimensional, while the space of necklaces, namely orderings on a circle, is $(n-1)!$ dimensional. Yet, $TCS_n$ is different.
The objective of this section is to characterize this representation by obtaining its character. In particular, the character will allow us to decompose it into irreducible components.

\presub {\bf Early attempts}. Initially we analyzed $TCS_n$ for low and specific $n$. We started with its original definition as trivalent trees up to Jacobi identities, see below (\ref{Jacobi}). We determined the characters of both the space of diagrams and the space of Jacobi identities. Then we decomposed both of them into irreps. In some cases we also found higher order relations: ``second order''  relations among the Jacobi relations, third order among the second order and so on. In this way the equations are divided into separate sectors corresponding to the the various irreps (which cannot mix). Even without reference to the actual form of the equations, one can conclude by counting in the usual way that given $V_\lam$ variables and $R_\lam$ relations for the irrep labelled by $\lam$, then the number of solutions (namely the dimension of this space) $S_\lam$ is bound by $\min \{0,V_\lam-R_\lam\} \le S_\lam \le V_\lam$, and in fact the lower bound will be saturated unless the relations are degenerate. 

Using this $S_n$ covariant analysis we obtained  probable candidates for the irrep content for $n=4,5,6$. Then we proceeded to confirm these candidates through a computerized and non-covariant analysis. That was done by assigning labels to the diagrams and then writing down each realization of the Jacobi identity. Clearly, this labeling of the diagrams renders the $S_n$ symmetry non-manifest, but it enables a computerized calculation of the rank of the Jacobi identities.

\subsection{Free Lie algebras and the Lie operad}

The original problem, that of capturing symmetries of the partial amplitudes which originate with those of the color structures, is now formulated mathematically as the problem of obtaining the $S_n$ character of the space of color structures. Happily it turns out that (at least at tree level) this problem was fully solved in the mathematics literature in \cite{GetzlerKapranov94}.\footnote{We are grateful to A. Khoroshkin for informing us of that.} In order to describe these results we shall first define and discuss certain relevant mathematical concepts.

\presub {\bf Free Lie algebras}. The Jacobi identity played a central role in our considerations, and we did not consider other algebraic relations which would be less general. The free Lie algebra over some set $A$, denoted by $L(A)$,  is exactly the Lie algebra generated by $A$ with no further relations apart  for antisymmetry and the Jacobi identity which are mandated by definition. Hence $L(A)$ is relevant here.  It has the property that any Lie algebra $L$ which contains $A$  defines a unique Lie homomorphism $L(A) \to L$. One of the early works employing free Lie algebras was \cite{Witt1937}. See the book \cite{Reutenauer} dedicated to free Lie algebras. For $|A| \ge 2$ $L(A)$ is infinite dimensional, yet we would be interested in a finite dimensional subspace known as its \emph{multilinear part}, defined by (Lie) words over $A$ where each letter appears exactly once. 

{\it Examples}. For two generators, namely $A_2=(1,2)$, the free Lie algebra has a single element with two letters, namely $[1,2]$, which is also the algebra's multi-linear part. Given 3 letters there are two linearly independent elements, namely $[1,[1,2]]$ and $[2,[1,2]]$. Proceeding in this manner, given the number of letters there are finitely many linearly independent elements, yet the total dimension of $L(A_2)$ is infinite. 

For 3 generators, namely $A_3=(1,2,3)$, the free Lie algebra has 3 linearly independent elements with two letters, namely $[1,2],\, [2,3]$ and $[3,1]$. The multi-linear part is here 3-lettered and is generated by $[[1,2],3]$ and $[[2,3],1]]$. The element $[[3,1],2]$ is linearly dependent on them due to the Jacobi identity.

\presub {\bf The Lie operad}. Informally, an operad can be thought to be the set of all possible expressions which can be generated from a given operation defined in some algebraic structure, see \cite{May1972} for the original definition.
One also defines cyclic operads. Informally these are operads where the inputs of the expression, namely its variables, are on equal footing with the output.  
For completeness we include the formal definition of an operad and a cyclic operad in appendix \ref{app:operads},  \footnote{
The general notion of a cyclic operad was introduced by \cite{GinzburgKapranov1994} following ideas of \cite{Kontsevich1994}.} 
 yet we find the examples below to be more illuminating. %

\newpage
{\it Examples} \bi

\item Commutative operad. 

This operad describes all the possible expressions that can be formed out of $n$ commuting elements, for instance $n$ commuting matrices. Clearly the expression is independent of the order, namely invariant under permutations, and therefore unique, and hence ${\it Com}(n)={\bf 1}$, where ${\bf 1}$ denotes the trivial representation of $S_n$. The associated cyclic operad can be realized by the trace of the product of $n$ commuting matrices and here too ${\it Com}((n))={\bf 1}$.

\item Associative operad.

The associative operad ${\it Assc}$ describes all the possible expressions that can be formed out of $n$ elements under an associative binary operation, for example $n$ general matrices. Here any ordering is independent and hence ${\it Assc}(n)$ has dimension $n!$ and in fact it is nothing but the regular representation. The cyclic version of this operad can be realized by the trace of a product of $n$ matrices. Due to the cyclic symmetry $\rm{dim}\({\it Assc}((n))\, \)=(n-1)!$ and mathematically it can described by ${\it Assc}((n))={\rm Ind}_{\IZ_n}^{S_n} {\bf 1}$, namely it is the representation of $S_n$ induced by the trivial representation of its $n$-cycle subgroup. The induced representation will be defined and further discussed below under  (\ref{reduce-Lie-cyc}).

\item Lie operad.

The Lie operad ${\cal L}ie (n)$ describes all the independent expressions that can be formed from $n$ elements by using the Lie bracket. It can be identified with the multilinear part of the free Lie algebra $L(A_n), ~A_n=(1 \dots n)$ 
and it can be realized as the set of rooted trees with oriented trivalent vertices, up to Jacobi identity, with leaves labeled by $1 \dots n$ denoting the ``input'' generators and a  root denoting the ``output'' generator. In these trees each vertex represents the Lie bracket operation. 

In the \emph{cyclic Lie operad} ${\cal L}ie((n))$ the trees are not rooted, but rather all external legs are of equal standing. As a set the cyclic ${\cal L}ie((n))$ is the same as the non-cyclic ${\cal L}ie(n-1)$, only ${\cal L}ie((n))$ transforms naturally under $S_n$, rather than $S_{n-1}$. The definition of the cyclic Lie operad uses the Killing form, namely the trace, in addition to the Lie bracket. 

For further detail see \cite{GetzlerKapranov94}, paragraphs 1.8.3-5, and the book \cite{operads} dedicated to operads and their various applications.
\ei

Comparing the above mentioned realization of $\Lie((n))$ through trees with the definition of $TCS_n$ through trees under (\ref{Jacobi}) 
 we recognize that \emph{the space of tree level color structures  is equivalent to the cyclic Lie operad}, namely \be
TCS_n \equiv {\cal L}ie((n)) ~.
\ee 
The cyclic property represents the equal standing of all external legs in a Feynman diagram.

\subsection{The nature of $TCS_n$}

The cyclic Lie operad, or equivalently $TCS_n$ can be described in terms of the non-cyclic one $\Lie$ by \be
TCS_n = {\rm Ind}_{S_{n-1}}^{S_n}\, \Lie(n-1) - \Lie(n)
\label{reduce-Lie-cyc}
\ee
which we now proceed to explain. The first term on the right hand side (RHS) denotes the $S_n$ representation induced by $\Lie(n-1)$ which is a representation of its $S_{n-1}$ subgroup. For completeness, we provide a concise definition of the general notion of an induced representation. Given a representation of $\pi:H \to V$, where $H$ is a subgroup of $G$, the representation induced onto $G$ is given by ${\rm Ind}_H^G \pi = K[G] \otimes_{K[H]} V$ where $K[G],K[H]$ are the group algebras. However, in this paper we shall not require this general definition, but rather we shall provide explicit descriptions of the relevant representations. Here ${\rm Ind}_{S_{n-1}}^{S_n} \Lie(n-1)$ is the set of all the expressions that can be formed by choosing $n-1$ out of the $n$ variables and using the Lie bracket operation to create expressions where each variables appears exactly once. The subtracted second term on the RHS implies that ${\rm Ind}_{S_{n-1}}^{S_n} \Lie(n-1)$ has a sub-representation which is equivalent to $\Lie(n)$ and is to be subtracted to obtain $TCS_n$. In other words $TCS_n + \Lie(n) = {\rm Ind}_{S_{n-1}}^{S_n}\, \Lie(n-1)$.

An equivalent form of this relation was shown to us by A. Khoroshkin, and while we do not know a reference where the proof could be found, we did subject it to numerous tests which the relation passed successfully. It is suggestive to interpret the relation as follows: for any $w \in TCS_n$ and for any choice of $1 \le j \le n$ there is a unique decomposition $w={\rm Tr} \( j w' \)$ where $w'$ belongs to $\Lie(n-1)$ on the set $1 \dots n$ with $j$ omitted and hence to ${\rm Ind}_{S_{n-1}}^{S_n}\, \Lie(n-1)$. Graphically this is equivalent to choosing one of the external legs to serve as a root. This decomposition is represented by the first term in  (\ref{reduce-Lie-cyc}). However, in this way we over-count $TCS_n$  and evidently the second term compensates for that.  

  
The non-cyclic operad $\Lie(n)$ can be described in turn by \cite{Klyachko1974} \be
\Lie(n) = {\rm Ind}_{Z_n}^{S_n}\, \om_n
\label{reduce-Lie-non-cyc}
\ee
which we now proceed to explain. $\om_n := \exp (2 \pi i/n )$ is a primitive root of unity. Here it represents the primitive representation of an $n$-cycle in $S_n$, namely we consider the subgroup $H$ of $S_n$ generated by the $n$-cycle $\rho_n=(1 2 \dots n) \to (2 \dots n 1)$ and its primitive representation is defined by $\rho_n \to \omega_n$.  This specific induced representation can be described more explicitly by the set of all orderings of $1 \dots n$ up to the equivalence relation  \be
\sig_1 \dots \sig_n \simeq \om_n\, \sig_2\,  \dots \sig_n\, \sig_1 
\ee
 where $\sig_1,\dots,\sig_n$ is any permutation of $(1 \dots n)$, and the equivalence means that a cyclic shift of an $n$-lettered word is allowed as long as it is accompanied by multiplication by the primitive root of unity.  See \cite{Reutenauer} theorem 8.3 for three proofs of (\ref{reduce-Lie-non-cyc}). In particular the third proof (based on \cite{Klyachko1974}) employs the notion of a Lie idempotent. 

Let us illustrate the expressions above by reproducing the dimension of $TCS_n$. From (\ref{reduce-Lie-non-cyc}) and the formula for the dimension of the induced representation, namely ${\rm dim}({\rm Ind}_H^G\, V) = |V|\, |G|/|H|$, we find ${\rm dim}\(\Lie(n)\)=1 \cdot n!/n=(n-1)!$. Now from (\ref{reduce-Lie-cyc}) we have ${\rm dim}(TCS_n) = (n-2)! n!/(n-1)!-(n-1)!=(n-2)!$ as we found in (\ref{dim}).

It is known that characters $\chi$ of the permutation (symmetric) group are conveniently given by the characteristic function of $\chi$, $ch(\chi)$, which is a symmetric function in some variables $x_i$. Being symmetric it can be expressed in terms of the power sums $p_k(x_i):=\sum_i x_i^k$, namely $ch=ch(p_1,p_2,\dots)$. The character is contained in the Taylor coefficients of the characteristic function. For a conjugacy class $c$ labeled by the partition of $n$ of the form $(1^{k_1} 2^{k_2} \dots)$ 
\footnote{This is the standard notation for a partition of $n$ into $k_1$ $1$s, $k_2$ $2$s and so on, such that $k_1 \cdot 1 + k_2 \cdot 2 + \dots = n$.}
its character is read through \be
\chi(c) = \( 1 \del_{p_1}\)^{k_1} \(2\, \del_{p_2}\)^{k_2} \dots ch(p_1,p_2,\dots) |_{0=p_1=p_2=\dots}
\label{chi}
\ee

In the non-cyclic case the characteristic functions are given by \cite{Brandt44,Thrall42} \be
ch_n \equiv ch(\Lie(n)) = \frac{1}{n} \sum_{d|n} \mu(d)\, p_d^{n/d}
\label{chn}
\ee 
where \be
\mu(d) := \left\{ \begin{array}{cc}
 0 		& \mbox{if $d$ is divisible by a square} \\
 (-)^k 	& \mbox{if $d$ is a product of $k$ distinct primes} 
 \end{array} \right.
 \ee
 is the M\"obius function from number theory.\footnote{
 We would like to record two important properties of the M\"obius function. First $\mu(n)=\sum_{(k,n)=1} \exp(2 \pi i k/n)$ where the sum is taken over all integers $k$ in the range $1 \le k \le n$ which are relatively prime to $n$. A second important property is $\sum_{d|n} \mu(d) = \delta_{n 1}$, which has a nice interpretation in terms of the Dirichlet convolution of number theory. For any two function of integers (or series) $f=f(n),\, g=g(n),\, n=1,2,\dots$ the Dirichlet convolution $f * g$ is defined by $\(f * g\)(n)= \sum_{k|n} f(k)\, g(n/k)$. The function $e(n):=\delta_{n1}$ 
 plays the role of a unit with respect to the Dirichlet convolution. Now the second property above can be rewritten as $\mu * 1 = e$ where $1$ is the constant function given by $1(n)=1$ for all $n$. This means that 
 $\mu$ is the inverse with respect to the Dirichlet convolution of the constant function $1$. Therefore if $f=\sum_{k|n} g(k)$, namely $f=g*1$ then $g=\mu * f$, namely $g(n)=\sum_{k|n} \mu(k) f(n/k)$ which is known as the M\"obius inversion formula.}
%
 
We note that the expression (\ref{chn}) is closely related to the generating function of primitive necklaces, see \cite{Reutenauer}.
 
 The characteristics (\ref{chn}) for all $n$ can be conveniently packaged (by formal summation) into a characteristic generating function for the whole operad \be
 ch(\Lie) := \sum_k ch_k = \sum_{n \ge 1} \frac{\mu(n)}{n}\, \log (1-p_n) ~.
 \label{ch}
 \ee
Formal expansion of the $\log$ into a power series reproduces (\ref{chn}).

Now the characteristic of the cyclic case can be stated. The generating function is \be
 Ch(TCS) = (p_1 -1)\, ch(\Lie) -p_1 
 \label{Ch}
 \ee
and the individual characteristics for $n \ge 3$ are\footnote{ 
We note a relative minus sign between our expressions and the original ones in \cite{GetzlerKapranov94}, presumably originating from a different convention.}
\be
 Ch_n \equiv  ch(TCS_n) = p_1\, ch_{n-1} - ch_n ~.
 \label{Chn}
 \ee

These characteristics of the cyclic Lie operad ($\equiv TCS$) were obtained in \cite{GetzlerKapranov94} through a certain Legendre transform on the commutative operad, which is rather simple to describe.\footnote{
The duality between commutative coalgebras and Lie algebras was observed independently by J. Moore and D. Quillen in the late 1960's according to \cite{operads} p.143.}
 Alternatively, they can be motivated by (\ref{reduce-Lie-cyc}): its first term implies the factor $p_1\, ch (\Lie)$ in (\ref{Ch}) while the second implies the term $-ch (\Lie)$. Note that the term $-p_1$ is inessential as it affects only $Ch_1$. A geometrical realization of (\ref{Ch}) was given in \cite{Getzler94}.

The expression (\ref{Ch}) is our main result (imported from \cite{GetzlerKapranov94}). From it one can read off the characteristic $Ch_n$ of all the tree-level color structures $TCS_n$, given by  (\ref{Chn}) where  (\ref{chn}) is used, and the character itself can be found with the help of (\ref{chi}). The character (and characteristic) are the sought-for characterization of the representation under permutations. If desired, the irrep components can now be found through multiplication with the character table. Explicit results for $n \le 9$ are listed in appendix \ref{sec:irrep}. Altogether this consists a full answer for the second issue mentioned in the introduction in the case of tree level.

Discussion:
\bi
\item Reducibility. In table \ref{tab:multirr} it is seen that $TCS_n$ is irreducible for $n=4,5$, but it is reducible for $n \ge 6$. For example $TCS_6$ decomposes as \bea
TCS_6 	& \to & (42) + (3 1^3) + (2^3) \\
4! 		& \to & 9 + 10 + 5
\eea
where the second line specifies the dimensions of the corresponding representations on the first line.

\item Self duality under Young conjugation. In table \ref{tab:multirr} we observe that for some values of $n$ $TSC_n$ is self-dual under Young conjugation, namely under the interchange of rows and columns in the Young diagrams, or equivalently under tensoring it with the sign representation. The self-dual values are $n=4,5,8, 9$, while $n=3, 6,7$ are not.

\item By inspection (at least for $n \le 9$) the following irreps do not appear in $TCS_n$: $(n), (n-1\, 1),(2 1^{n-2})$ and $(1^n)$ (the first is the symmetric and the last is the anti-symmetric).

\item Interpretation for (\ref{reduce-Lie-cyc}). The induction in this formula translates in a standard way to the addition of a box in all possible ways to the irreps composing the inducing representation $\Lie(n-1)$. 

\item Support of the character. The result (\ref{Ch}) implies that the character is non-zero only for classes labeled by partitions of the form $(d^{n/d})$, namely partitions of $n$ into equal parts, or $(1 d^{(n-1)/d})$, which are partitions of $n-1$ into equal parts. 

\item For $n \le 6$ we found that the results in the table agree with those found through the early attempts described at the beginning of this section.

\ei

\section{Implications for sub-amplitudes}
\label{sec:implications}

The ultimate objective for studying the representation of the color structures is to assist the determination of partial amplitudes by separating their implied symmetry structure. Here we take a first and partial step in this direction, by showing a sort of converse. Namely, we show that given a set of objects $(ij)$ $1 \le i,j \le n$ satisfying 
\begin{gather}
(ij)=-(ji) \label{as} \\
(ij)(kl)+(il)(jk)+(ik)(lj)=0 \label{schouten} 
\end{gather}
then the following expressions
\begin{equation}
C_n(12...n):=\frac{1}{(12)(23)...(n1)} \label{cn}
\end{equation}
satisfy the Kleiss-Kuijf relations and hence all the symmetries of color-ordered amplitudes which are implied by the color structure symmetries.

{\bf Motivation}. We notice that the permutation dependent factor in both the Parke-Taylor expression \cite{ParkeTaylor1986} and the Cachazo-He-Yuan (CHY) formula \cite{CachazoHeYuan2013} are of this form. In Parke-Taylor the permutation-dependent factor is $\[ \langle 12 \rangle \langle 23 \rangle ... \langle n1 \rangle \]^{-1}$ and the spinor products $(ij) := \langle i\, j \rangle$ satisfy (\ref{as},\ref{schouten}). In CHY the permutation-dependent factor is \begin{equation}
\frac{1}{\sigma_{12} \sigma_{23} ... \sigma_{n1}}
\end{equation}
where $\sigma_i \in \IC$ is a set of solutions of the scattering (Gross-Mende) equations, and $(ij) := \sigma_i-\sigma_j$ satisfy (\ref{as},\ref{schouten}) too. We note that the relations (\ref{as},\ref{schouten}) can be interpreted geometrically to mean that $(ij)$ belongs to the Grassmannian $Gr(2,n)$, the space of 2-planes in $n$-dimensional space. Indeed the first identity implies that $\om_{ij}:=(ij)$ is a bivector, while the second implies that $\om \wedge \om =0$ and hence $\om = u \wedge v$ for some vectors $u,v$ and it represents a 2-plane.

Note that a claim equivalent to the one proven here was made in \cite{TyeZhang2010b}, and a general proof for any $n$ which was lacking there is given here. It appears that essentially the same argument was given in  \cite{BernDixonDunbarKosower94}, see eq. (III.9), in a somewhat different context, that of sub-leading color factors for 1-loop amplitudes. Our approach differs by being axiomatic and hence applying immediately not only to Parke-Taylor amplitudes but also to CHY.

\presub {\bf Proof}. We will prove inductively on $q \equiv |\beta|$ the Kleiss-Kuijf relations 
\begin{equation}
C_n(1\, \beta_1 \beta_2 ... \beta_q \, 2 \, \alpha_1 \alpha_2 ... \alpha_p)=(-1)^{q} \sum_{\sigma \in \alpha_1 \alpha_2 ... \alpha_p \shuffle \beta_q \beta _{q-1}...\beta _1} C_n(1\,2\, \sigma). 
\label{th:kk}
\end{equation}
Without loss of generality we will assume $q \leq p$ (otherwise we can interchange the roles of $\alpha$ and $\beta$ and use the reflectivity of $C_n$). 

For the base case of $q=1$ of (\ref{th:kk}) we want to prove that 
\begin{equation}
0 =\frac{1}{(1\,2)(2\alpha_1)...(\alpha_p1)}\Big[\frac{(1\,2)}{(1 \beta_1)(\beta_1 2)}+\sum_{k=0}^p \frac{(\alpha_k \alpha_{k+1})}{(\alpha_k \beta_1)(\beta_1 \alpha_{k+1})}\Big] \label{kk1}
\end{equation}
with the convention $\alpha_0=2$ and $\alpha_{p+1}=1$. Multiplying the term in the sum by $\frac{(\beta_1 2)}{(\beta_1 2)}$ and using (\ref{schouten}) we get 
\begin{equation}
\frac{1}{(\beta_1 2)} \sum_{k=0}^p \Big(\frac{(\alpha_k 2)}{(\alpha_k \beta_1)}-\frac{(\alpha_{k+1} 2)}{(\alpha_{k+1} \beta_1)}\Big)=\frac{1}{(\beta_1 2)}\Big(\frac{(2\,2)}{(2\beta_1)}-\frac{(1\,2)}{(1\beta_1)}\Big)=-\frac{(1\,2)}{(1\beta_1)(\beta_1 2)}
\end{equation}
which cancels exactly the first term in the parenthesis in (\ref{kk1}).

For the inductive step we assume (\ref{th:kk}) for $|\beta|=q$ and rewrite the shuffle more explicitly as 
\begin{eqnarray}
0 &=& \frac{1}{(1\,2)(2\alpha_1)...(\alpha_p1)}\Big[\frac{(1\,2)}{(1 \beta_1)...(\beta_q 2)} \nonumber \\
&-&(-1)^q \Big( \sum_{k=0}^p \frac{(\alpha_k \alpha_{k+1})}{(\alpha_k \, \beta \, \alpha_{k+1})}
+\sum_{\beta^T= \beta ' \beta ''} \sum_{k_1=1}^p \sum_{k_0=0}^{k_1-1} \frac{(\alpha_{k_0} \alpha_{k_0+1})}{(\alpha_{k_0} \beta ' \alpha_{k_0+1})}\frac{(\alpha_{k_1} \alpha_{k_1+1})}{(\alpha_{k_1} \beta '' \alpha_{k_1+1})} \nonumber \\
&+&\sum_{\beta^T=\beta ' \beta '' \beta '''} \sum_{k_2=2}^p \sum_{k_1=1}^{k_2-1} \sum_{k_0=0}^{k_1-1} \frac{(\alpha_{k_0} \alpha_{k_0+1})}{(\alpha_{k_0} \beta ' \alpha_{k_0+1})}\frac{(\alpha_{k_1} \alpha_{k_1+1})}{(\alpha_{k_1} \beta '' \alpha_{k_1+1})} \frac{(\alpha_{k_2} \alpha_{k_2+1})}{(\alpha_{k_2} \beta ''' \alpha_{k_2+1})} \nonumber \\
&+&...+
\sum_{k_{q-1}=q-1}^p ...\sum_{k_0=0}^{k_1-1}  \frac{(\alpha_{k_0} \alpha_{k_0+1})}{(\alpha_{k_0} \beta_q) (\beta_q \alpha_{k_0+1})}... \frac{(\alpha_{k_{q-1}} \alpha_{k_{q-1}+1})}{(\alpha_{k_{q-1}} \beta_1)(\beta_1 \alpha_{k_{q-1}+1})} \Big) \Big] \label{kkq}
\end{eqnarray}
where $\sum_{\beta^T =\beta ' \beta ''...\beta^{(n)}}$  means the sum over all partitions of the sequence $\beta^T \equiv \beta_q \dots \beta_1$ into $n$ parts, and $(\alpha_k \, \beta \, \alpha_{k+1})$ is shorthand for $(\alpha_k \beta_q)(\beta_q \beta_{q-1})...(\beta_1 \alpha_{k+1})$. 

Now we shall prove this for $|\beta|=q+1$ using (\ref{kkq}) as our induction assumption. When we add $\beta_{q+1}$ to the terms in (\ref{kkq}) we can do so in two ways: either attach $\beta_{q+1}$ to $\beta_q$ in the same partition, or add a new partition containing only $\beta_{q+1}$. In this way a typical term in (\ref{kkq}), say
\begin{eqnarray}
\sum_{\beta^T = \beta ' \beta '' \beta '''} \sum_{k_2=2}^p \sum_{k_1=1}^{k_2-1} \sum_{k_0=0}^{k_1-1} \frac{(\alpha_{k_0} \alpha_{k_0+1})}{(\alpha_{k_0} \beta ' \alpha_{k_0+1})}\frac{(\alpha_{k_1} \alpha_{k_1+1})}{(\alpha_{k_1} \beta '' \alpha_{k_1+1})} \frac{(\alpha_{k_2} \alpha_{k_2+1})}{(\alpha_{k_2} \beta ''' \alpha_{k_2+1})} \label{typq}
\end{eqnarray}
becomes the following two terms
\begin{eqnarray}
& & \sum_{\beta^T = \beta ' \beta ''\beta '''} \sum_{k_2=2}^p \sum_{k_1=1}^{k_2-1} \sum_{k_0=0}^{k_1-1} \frac{(\alpha_{k_0} \alpha_{k_0+1})}{(\alpha_{k_0} \beta_{q+1} \beta ' \alpha_{k_0+1})}\frac{(\alpha_{k_1} \alpha_{k_1+1})}{(\alpha_{k_1} \beta '' \alpha_{k_1+1})} \frac{(\alpha_{k_2} \alpha_{k_2+1})}{(\alpha_{k_2} \beta ''' \alpha_{k_2+1})} \label{typ} \\
&+& \sum_{\beta^T = \beta ' \beta '' \beta '''} \sum_{k_3=3}^p \sum_{k_2=2}^{k_3-1} \sum_{k_1=1}^{k_2-1} \sum_{k_0=0}^{k_1-1} \frac{(\alpha_{k_0} \alpha_{k_0+1})}{(\alpha_{k_0} \beta_{q+1} \alpha_{k_0+1})}\frac{(\alpha_{k_1} \alpha_{k_1+1})}{(\alpha_{k_1} \beta ' \alpha_{k_1+1})} \frac{(\alpha_{k_2} \alpha_{k_2+1})}{(\alpha_{k_2} \beta '' \alpha_{k_2+1})}  \frac{(\alpha_{k_3} \alpha_{k_3+1})}{(\alpha_{k_3} \beta ''' \alpha_{k_3+1})}~. \nonumber 
\end{eqnarray}

Multiplying the second term in (\ref{typ}) by $\frac{(\beta_{q+1}\beta_q)}{(\beta_{q+1}\beta_q)}$ and using (\ref{schouten}) over $(\beta_{q+1}\beta_q)(\alpha_{k_0} \alpha_{k_0+1})$ we get for this term

\begin{eqnarray}
\frac{1}{(\beta_{q+1}\beta_q)} \sum_{k_3=3}^p \sum_{k_2=2}^{k_3-1} \sum_{k_1=1}^{k_2-1} \sum_{k_0=0}^{k_1-1} \Big(\frac{(\alpha_{k_0} \beta_q)}{(\alpha_{k_0} \beta_{q+1})}-\frac{(\alpha_{k_0+1} \beta_q)}{(\alpha_{k_0+1} \beta_{q+1})}\Big) \nonumber \\
\times \frac{(\alpha_{k_1} \alpha_{k_1+1})}{(\alpha_{k_1} \beta ' \alpha_{k_1+1})} \frac{(\alpha_{k_2} \alpha_{k_2+1})}{(\alpha_{k_2} \beta '' \alpha_{k_2+1})}  \frac{(\alpha_{k_3} \alpha_{k_3+1})}{(\alpha_{k_3} \beta ''' \alpha_{k_3+1})}~. \label{typ1}
\end{eqnarray}

Performing the telescopic sum over $k_0$ in (\ref{typ1}), we get from the second term in (\ref{typ}) two terms
\begin{eqnarray}
& &\frac{1}{(\beta_{q+1}\beta_q)} \frac{(2 \beta_q)}{(2 \beta_{q+1})}\sum_{\beta^T = \beta ' \beta '' \beta '''} \sum_{k_2=2}^p \sum_{k_1=1}^{k_2-1} \sum_{k_0=0}^{k_1-1} \frac{(\alpha_{k_0} \alpha_{k_0+1})}{(\alpha_{k_0} \beta ' \alpha_{k_0+1})}\frac{(\alpha_{k_1} \alpha_{k_1+1})}{(\alpha_{k_1} \beta '' \alpha_{k_1+1})} \frac{(\alpha_{k_2} \alpha_{k_2+1})}{(\alpha_{k_2} \beta ''' \alpha_{k_2+1})} \nonumber\\
&-& \sum_{\beta^T = \beta ' \beta '' \beta '''} \sum_{k_2=2}^p \sum_{k_1=1}^{k_2-1} \sum_{k_0=0}^{k_1-1} \frac{(\alpha_{k_0} \alpha_{k_0+1})}{(\alpha_{k_0} \beta_{q+1} \beta ' \alpha_{k_0+1})}\frac{(\alpha_{k_1} \alpha_{k_1+1})}{(\alpha_{k_1} \beta '' \alpha_{k_1+1})} \frac{(\alpha_{k_2} \alpha_{k_2+1})}{(\alpha_{k_2} \beta ''' \alpha_{k_2+1})}~. \label{typ2}
\end{eqnarray}
We notice two things: first, the last term in (\ref{typ2}) exactly cancels the first term in (\ref{typ}); and second, the first term in (\ref{typ2}) is $\frac{1}{(\beta_{q+1}\beta_q)} \frac{(2 \beta_q)}{(2 \beta_{q+1})}$ times the term we've begun with in (\ref{typq}) for $|\beta|=q$. Therefore for $|\beta|=q+1$ we are left with
\begin{eqnarray}
\frac{(1\,2)}{(1 \beta_1)...(\beta_q\beta_{q+1})(\beta_{q+1} 2)}
-(-1)^{q+1} \frac{1}{(\beta_{q+1}\beta_q)} \frac{(2 \beta_q)}{(2 \beta_{q+1})} \Big( \text{terms from $|\beta|=q$}\Big)~.
\end{eqnarray}
Taking out $\frac{1}{(\beta_{q+1}\beta_q)} \frac{(2 \beta_q)}{(2 \beta_{q+1})}$ as a common factor we use the induction assumption (\ref{kkq}) to prove this is zero. 

We therefore managed to prove directly that objects of the form (\ref{cn}), where $(ij)$ is any collection of objects satisfying (\ref{as}) and (\ref{schouten}), satisfy the Kleiss-Kuijf relations (\ref{th:kk}).

\section{Conclusions}
\label{sec:conclude}

In section \ref{sec:color} we reviewed how the color-ordered sub-amplitudes is a set of components of the total amplitude with respect to a set of bases all of equal standing, and this is summarized in subsection \ref{subsec:interim-summ}.

Our main results are \bi
	\item The mathematical formulation for symmetries of the partial amplitudes which originate with those of color structures, see the beginning of section \ref{sec:permute} and earlier.
	
	 \item The identification of the space of tree-level color structure with the Lie operad and its characterization (as a representation of the permutation group) by the generating function (\ref{Ch}). It is imported from the mathematical literature \cite{GetzlerKapranov94}, and we are unaware of an earlier linkage between the two. 
	 
	 \item The explicit list of permutation irreps for $n \le 9$ in table  \ref{tab:multirr}.
\ei 
We believe that this determination of symmetries should be useful for the determination of partial amplitudes. We took a first step in this direction in section \ref{sec:implications} where we proved that the permutation-dependent part of both the Parke-Taylor amplitudes and the Cachazo-He-Yuan amplitudes can be shown to satisfy the Kleiss-Kuijf relations just by  using certain Grassmannian relations. Other implications are left for future work.
 
Another direction for future work is to obtain a simplified expression for the irrep multiplicities.

\subsection*{Acknowledgments}

It is a pleasure to thank R. Adin,  E. Getzler, A. Hanany, Y. Roichman, Z. Sela  and especially A. Khoroshkin for discussions on mathematical aspects and members of the high energy group at the Hebrew University of Jerusalem for comments on a presentation.  We are thankful to L. Dixon for a discussion and very useful comments on a draft.
BK is grateful to the organizers of the Kallosh / Shenker fest at Stanford and to J. Maldacena and N. Arkani-Hamed for hospitality at the Princeton Institute for Advanced Study during the initial stages of this work. 

This research was supported by the Israel Science Foundation grant no. 812/11 and it is part of the Einstein Research Project "Gravitation and High Energy Physics", which is funded by the Einstein Foundation Berlin.

\appendix

\section{Definition of Operads}
\label{app:operads}

In this appendix we provide a formal definition of operads closely following the presentation of \cite{GetzlerKapranov94}, section 1. 

A chain complex (or dg-vector space) is a graded vector space $U_\bullet$ together with a differential
$\delta: U_i \to U_{i-1}$, such that $\delta^2$ = 0.

An $\IS$-module is a sequence of chain complexes ${\cal U} = \{ {\cal U}(n) | n \ge 0 \}$, together with an action of the symmetric group $S_n$ on ${\cal U}(n)$ for each $n$.

An operad is an $\IS$-module $Q$ together with bilinear operations \be
\circ_i : Q(m) \otimes Q(n) \to Q(m + n - 1), ~~ 1 \le i \le m, \ee
satisfying the following axioms.

\begin{enumerate}

\item If $\sigma \in S_m, \rho \in S_n, a \in Q(m)$ and $b \in Q(n)$ then \be
(\sigma a) \circ_{\sigma(i)} (\rho b) = ( \sigma \circ_i \rho) ( a \circ_i b) ~,
\ee
where $\sigma \circ_i \rho \in S_{n+1}$ is defined to permute the interval $\{ i, \dots ,i+n-1 \}$ according to the permeation $\rho$ and then reorders the $m$ intervals \be
\{1\},\dots,\{i-1\}, \{ i, \dots ,i+n-1 \}, \{i+n \}, \dots, \{ m+n-1 \} ~,
\ee
which partition $\{1, \dots,m+n-1\}$ according to $\sigma$. Explicitly \be
 \( \sigma \circ_i \rho \) (j) = \left\{ \begin{array}{ll} 
	\sigma(j) & \qquad  j < i \mbox{ and } \sigma(j)<\sigma(i)  \\
	\sigma(j) + n-1 & \qquad j<i \mbox{ and } \sigma(j)>\sigma(i) \\
	\sigma(j-n+1) & \qquad  j \ge  i+n \mbox{ and } \sigma(j)<\sigma(i) \\
	\sigma(j-n+1)+n-1 & \qquad  j \ge  i+n \mbox{ and } \sigma(j)>\sigma(i) \\
	\sigma(i) + \rho(j-i+1) -1 & \qquad  i \le j  < i + n  
\end{array} \right.
\ee

\item For any $a \in Q(k), b \in Q(l)$ and $c \in Q(m)$ and any $1 \le i < j  \le k$, \be
 \( a \circ_i b \) \circ_{i+j-1} c = a \circ_i \( b \circ_j c\).
\ee 

\item For any $a \in Q(k), b \in Q(l)$ and $c \in Q(m)$ and any $1 \le i  \le k, 1 \le j  \le l$, \be
 \( a \circ_i b \) \circ_{i+j-1} c = a \circ_i \( b \circ_j c \) .
 \ee

\end{enumerate}

Realization through trees. We may associate with an element of $Q(n)$  a rooted
tree with $n$ inputs numbered from $1$ up to $n$ and one output. The compositions translate into grafting two such trees together along the $i$th input of the first tree.

\subsubsection*{Cyclic operads} 

A cyclic $\IS$-module ${\cal U}$ is a sequence of vector spaces ${\cal U}(n)$ each acted upon by $S_{n+}$, defined to be the group of permutations on $0 1 \dots n$.

For a cyclic $\IS$-module ${\cal U}$ and a $(k+1)$-element set $I$ define \be
{\cal U} ((I)) = \( \bigoplus_{ \mbox{bijections } f:\{0,\dots,k\} \to I } {\cal U}(k) \)_{S_{k+}}
\ee
In the case when $k=n-1$ and $I=\{1,\dots,n\}$ we write ${\cal U}((n))$ instead of ${\cal U}((I))$ and we note that ${\cal U}((n)) = {\cal U}(n-1)$, which means that as a vector space a cyclic operad with $n$ inputs can be identified with a non-cyclic operad with $n-1$ inputs after one of the original inputs was arbitrarily designated as output.

A cyclic operad is a cyclic $\IS$-module $Q$ which satisfies the following condition.  For any $a \in Q(n)$ let $a^*$ be the result of applying the cycle $(0 1\dots n) \in S_{n+}$ to $a$. Now the condition is given by \be
\( a \circ_n b \)^* = b^* \circ_1 a^* ~,
\ee 
for any $a \in Q(n),\, b \in Q(m)$. This means that permutations are allowed to permute the output of the operad with its inputs.

For further information and discussion see \cite{GetzlerKapranov94}.

\section{Irreducible $S_n$ components}
\label{sec:irrep}

Table \ref{tab:multirr} on p.25 shows the multiplicities of the irreducible representation of $S_n$ for $\Lie(n)$ and $\Lie((n)) \equiv TCS_n$. The non-cyclic case was obtained in \cite{Thrall42,Brandt44} and is reproduced here for convenience. \cite{GAP4} was used for computerized calculations.

In the almost trivial case of $n=3$ we have $\Lie(3)=(21)$ and $\Lie((3))=(1^3)$.

\begin{table}[t!]
\caption{Multiplicities of the irreducible representation of $S_n$ for $\Lie(n)$ and $\Lie((n)) \equiv TCS_n$. The notation $\lambda=a^i b^j...$ describes a Young diagram with $a$ boxes in each of the first $i$ rows and $b$ boxes in each of the next $j$ rows, etc. or equivalently a partition of $n$, $i \cdot a + j \cdot b + \dots = n$.}
\centering
\begin{tabular}{c c c c | c c c c}
\hline
$n$ & $\lambda$ & $\Lie(n)$ & $TCS_n$ & $n$ & $\lambda$ & $\Lie(n)$ & $TCS_n$\\
\hline
4 & 31 & 1 & 0 & & $421^2$ & 12 & 1\\
  & $2^2$ & 0 & 1 & & $41^4$ & 4 & 1\\
  & $21^2$ & 1 & 0 & & $3^22$ & 6 & 0\\
5 & 41 & 1 & 0 & & $3^21^2$ & 6 & 2\\
  & 32 & 1 & 0 & & $32^21$ & 9 & 1\\
  & $31^2$ & 1 & 1 & & $321^3$ & 8 & 1\\
  & $2^21$ & 1 & 0 & & $31^5$ & 3 & 0\\
  & $21^3$ & 1 & 0 & & $2^4$ & 1 & 1\\
6 & 51 & 1 & 0 & & $2^31^2$ & 4 & 0\\
  & 42 & 1 & 1 & & $2^21^4$ & 2 & 1\\
  & $41^2$ & 2 & 0 & & $21^6$ & 1 & 0\\
  & $3^2$ & 1 & 0 & 9 & 81 & 1 & 0\\
  & 321 & 3 & 0 & & 72 & 3 & 0\\
  & $31^3$ & 1 & 1 & & $71^2$ & 3 & 1\\
  & $2^3$ & 0 & 1 & & 63 & 5 & 1\\
  & $2^21^2$ & 2 & 0 & & 621 & 12 & 1\\
  & $21^4$ & 1 & 0 & & $61^3$ & 6 & 1\\
7 & 61 & 1 & 0 & & 54 & 5 & 0\\
  & 52 & 2 & 0 & & 531 & 18 & 3\\
  & $51^2$ & 2 & 1 & & $52^2$ & 13 & 1\\
  & 43 & 2 & 0 & & $521^2$ & 21 & 3\\
  & 421 & 5 & 1 & & $51^4$ & 8 & 0\\
  & $41^3$ & 3 & 0 & & $4^21$ & 9 & 1\\
  & $3^21$ & 3 & 1 & & 432 & 19 & 2\\
  & $32^2$ & 3 & 0 & & $431^2$ & 24 & 3\\
  & $321^2$ & 5 & 1 & & $42^21$ & 24 & 3\\
  & $31^4$ & 2 & 0 & & $421^3$ & 21 & 3\\
  & $2^31$ & 2 & 0 & & $41^5$ & 6 & 1\\
  & $2^21^3$ & 2 & 1 & & $3^3$ & 4 & 2\\
  & $21^5$ & 1 & 0 & & $3^221$ & 19 & 2\\
8 & 71 & 1 & 0 & & $3^21^3$ & 13 & 1\\
  & 62 & 2 & 1 & & $32^3$ & 9 & 1\\
  & $61^2$ & 3 & 0 & & $32^21^2$ & 18 & 3\\
  & 53 & 4 & 0 & & $321^4$ & 12 & 1\\
  & 521 & 8 & 1 & & $31^6$ & 3 & 1\\
  & $51^3$ & 4 & 1 & & $2^41$ & 5 & 0\\
  & $4^2$ & 1 & 1 & & $2^31^3$ & 5 & 1\\
  & 431 & 9 & 1 & & $2^21^5$ & 3 & 0\\
  & $42^2$ & 6 & 2 & & $21^7$ & 1 & 0\\
\hline
\end{tabular}
\label{tab:multirr} 
\end{table}

\bibliographystyle{unsrt}

\begin{thebibliography}{99}

\bibitem{DixonRev2011} 
  L.~J.~Dixon,
 ``Scattering amplitudes: the most perfect microscopic structures in the universe,''
  J.\ Phys.\ A {\bf 44}, 454001 (2011)
  arXiv:1105.0771 [hep-th].

\bibitem{Witten2003} 
  E.~Witten,
  ``Perturbative gauge theory as a string theory in twistor space,''
  Commun.\ Math.\ Phys.\  {\bf 252}, 189 (2004)
  [hep-th/0312171].

\bibitem{ElvangHuang2013} 
  H.~Elvang and Y.~-t.~Huang,
  ``Scattering Amplitudes,''
  arXiv:1308.1697 [hep-th].

\bibitem{DeWitt1967} 
  B.~S.~DeWitt,
  ``Quantum Theory of Gravity. 3. Applications of the Covariant Theory,''
  Phys.\ Rev.\  {\bf 162}, 1239 (1967).

\bibitem{ManganoParke1990} 
  M.~L.~Mangano and S.~J.~Parke,
  ``Multiparton amplitudes in gauge theories,''
  Phys.\ Rept.\  {\bf 200}, 301 (1991)
  [hep-th/0509223].



\bibitem{ParkeTaylor1986} 
  S.~J.~Parke and T.~R.~Taylor,
  ``An Amplitude for $n$ Gluon Scattering,''
  Phys.\ Rev.\ Lett.\  {\bf 56}, 2459 (1986).

\bibitem{BerendsGiele1987b} 
  F.~A.~Berends and W.~T.~Giele,
  ``Recursive Calculations for Processes with n Gluons,''
  Nucl.\ Phys.\ B {\bf 306}, 759 (1988).

\bibitem{Zeppenfeld88} 
  D.~Zeppenfeld,
  ``Diagonalization of Color Factors,''
  Int.\ J.\ Mod.\ Phys.\ A {\bf 3}, 2175 (1988).

\bibitem{BernDixonKosower1996} 
  Z.~Bern, L.~J.~Dixon and D.~A.~Kosower,
  ``Progress in one loop QCD computations,''
  Ann.\ Rev.\ Nucl.\ Part.\ Sci.\  {\bf 46}, 109 (1996)
  [hep-ph/9602280].

\bibitem{PenroseTwistors1967} 
  R.~Penrose,
  ``Twistor algebra,''
  J.\ Math.\ Phys.\  {\bf 8}, 345 (1967).

\bibitem{BCFW2005} 
  R.~Britto, F.~Cachazo, B.~Feng and E.~Witten,
  ``Direct proof of tree-level recursion relation in Yang-Mills theory,''
  Phys.\ Rev.\ Lett.\  {\bf 94}, 181602 (2005)
  [hep-th/0501052].

\bibitem{BCJ2008} 
  Z.~Bern, J.~J.~M.~Carrasco and H.~Johansson,
  Phys.\ Rev.\ D {\bf 78}, 085011 (2008)
  arXiv:0805.3993 [hep-ph].

\bibitem{ArkaniHamedCachazoCheung2009} 
  N.~Arkani-Hamed, F.~Cachazo and C.~Cheung,
  ``The Grassmannian Origin Of Dual Superconformal Invariance,''
  JHEP {\bf 1003}, 036 (2010)
  arXiv:0909.0483 [hep-th].

\bibitem{PositiveGrassmannian2012} 
  N.~Arkani-Hamed, J.~L.~Bourjaily, F.~Cachazo, A.~B.~Goncharov, A.~Postnikov and J.~Trnka,
  ``Scattering Amplitudes and the Positive Grassmannian,''
  arXiv:1212.5605 [hep-th].

\bibitem{Amplituhedron2013} 
  N.~Arkani-Hamed and J.~Trnka,
  ``The Amplituhedron,''
  arXiv:1312.2007 [hep-th].


\bibitem{Hodges2012} 
  A.~Hodges,
  ``A simple formula for gravitational MHV amplitudes,''
  arXiv:1204.1930 [hep-th].


\bibitem{CachazoHeYuan2013} 
  F.~Cachazo, S.~He and E.~Y.~Yuan,
  ``Scattering of Massless Particles in Arbitrary Dimension,''
  arXiv:1307.2199 [hep-th].


\bibitem{DixonRev2013} 
  L.~J.~Dixon,
  ``A brief introduction to modern amplitude methods,''
  arXiv:1310.5353 [hep-ph].

\bibitem{KeppelerSjodajl12} 
  S.~Keppeler and M.~Sjodahl,
  ``Orthogonal multiplet bases in SU(Nc) color space,''
  JHEP {\bf 1209}, 124 (2012)
  arXiv:1207.0609 [hep-ph].

\bibitem{EdisonNaculich12} 
  A.~C.~Edison and S.~G.~Naculich,
  ``Symmetric-group decomposition of SU(N) group-theory constraints on four-, five-, and six-point color-ordered amplitudes,''
  JHEP {\bf 1209}, 069 (2012)
  arXiv:1207.5511 [hep-th].

  \bibitem{Kanning:2014maa} 
  N.~Kanning, T.~Lukowski and M.~Staudacher,
  ``A shortcut to general tree-level scattering amplitudes in $\mathcal{N} = 4$ SYM via integrability,''
  Fortsch.\ Phys.\  {\bf 62}, 556 (2014)
  arXiv:1403.3382 [hep-th].
  
 \bibitem{Broedel:2014pia} 
  J.~Broedel, M.~de Leeuw and M.~Rosso,
  ``A dictionary between R-operators, on-shell graphs and Yangian algebras,''
  JHEP {\bf 1406}, 170 (2014)
  arXiv:1403.3670 [hep-th].

\bibitem{KleissKuijf1988} 
  R.~Kleiss and H.~Kuijf,
  ``Multi - Gluon Cross-sections and Five Jet Production at Hadron Colliders,''
  Nucl.\ Phys.\ B {\bf 312}, 616 (1989).

\bibitem{ReuschleWeinzierl13} 
  C.~Reuschle and S.~Weinzierl,
  ``Decomposition of one-loop QCD amplitudes into primitive amplitudes based on shuffle relations,''
  Phys.\ Rev.\ D {\bf 88}, 105020 (2013)
  arXiv:1310.0413 [hep-ph].

\bibitem{Cvitanovic-etal1981} 
  P.~Cvitanovic, P.~G.~Lauwers and P.~N.~Scharbach,
  ``Gauge Invariance Structure of Quantum Chromodynamics,''
  Nucl.\ Phys.\ B {\bf 186}, 165 (1981).

\bibitem{DelDucaDixon-etal1999a}
  V.~Del Duca, A.~Frizzo and F.~Maltoni,
  ``Factorization of tree QCD amplitudes in the high-energy limit and in the collinear limit,''
  Nucl.\ Phys.\ B {\bf 568}, 211 (2000)
  [hep-ph/9909464]. 

\bibitem{DelDucaDixon-etal1999b} 
  V.~Del Duca, L.~J.~Dixon and F.~Maltoni,
  ``New color decompositions for gauge amplitudes at tree and loop level,''
  Nucl.\ Phys.\ B {\bf 571}, 51 (2000)
  [hep-ph/9910563].


\bibitem{BerendsGiele1987a} 
  F.~A.~Berends and W.~Giele,
  ``The Six Gluon Process as an Example of Weyl-Van Der Waerden Spinor Calculus,''
  Nucl.\ Phys.\ B {\bf 294}, 700 (1987).


 \bibitem{ManganoParkeXu1987} 
  M.~L.~Mangano, S.~J.~Parke and Z.~Xu,
  ``Duality and Multi - Gluon Scattering,''
  Nucl.\ Phys.\ B {\bf 298}, 653 (1988).

\bibitem{Mangano1988} 
  M.~L.~Mangano,
  ``The Color Structure of Gluon Emission,''
  Nucl.\ Phys.\ B {\bf 309}, 461 (1988).

\bibitem{ChanPaton1969} 
  J.~E.~Paton and H.~-M.~Chan,
  ``Generalized veneziano model with isospin,''
  Nucl.\ Phys.\ B {\bf 10}, 516 (1969).


\bibitem{PeskinSchroeder}
M.~E.~Peskin and D.~V.~Schroeder,
``An Introduction to Quantum Field Theory,''
{\it Westview Press} 1995.

\bibitem{BjerrumBohr:2012mg} 
  N.~E.~J.~Bjerrum-Bohr, P.~H.~Damgaard, R.~Monteiro and D.~O'Connell,
  ``Algebras for Amplitudes,''
  JHEP {\bf 1206}, 061 (2012)
  arXiv:1203.0944 [hep-th].

\bibitem{Cvitanovic76} 
  P.~Cvitanovic,
  ``Group theory for Feynman diagrams in non-Abelian gauge theories,''
  Phys.\ Rev.\ D {\bf 14}, 1536 (1976).

\bibitem{BarNathan95}
D.~Bar-Natan, 
``Weights of Feynman diagrams and the Vassiliev knot invariants,''
www.math.toronto.edu/$\sim$drorbn/LOP.html (unpublished preprint, 1991); ``On the
Vassiliev knot invariants,'' Topology {\bf 34}, 423 (1995).

\bibitem{CvitanovicGroupTheory08} 
  P.~Cvitanovic,
  ``Group theory: Birdtracks, Lie's and exceptional groups,''
  Princeton, USA: Univ. Pr. (2008) 273 p.

\bibitem{DiracQM}
 P.~A.~M.~Dirac,
 ``The principles of quantum mechanics,'' {\it Oxford University Press} 1930.

\bibitem{Reutenauer}
  C.~Reutenauer, 
  ``Free Lie algebras,'' {\it The Clarendon Press, Oxford University Press}
  New York, 1993.

\bibitem{BjerrumBohrDamgaardVanhove09} 
  N.~E.~J.~Bjerrum-Bohr, P.~H.~Damgaard and P.~Vanhove,
  ``Minimal Basis for Gauge Theory Amplitudes,''
  Phys.\ Rev.\ Lett.\  {\bf 103}, 161602 (2009)
  arXiv:0907.1425 [hep-th]. \\
  S.~Stieberger,
  ``Open \& Closed vs. Pure Open String Disk Amplitudes,''
  arXiv:0907.2211 [hep-th].

\bibitem{GetzlerKapranov94}
 E.~Getzler and M.~Kapranov,
 ``Modular operads," Compositio Math. {\bf 110} 65 (1998) 
 [dg-ga/9408003]. 

\bibitem {Witt1937}
E.~Witt,
``Treue Darstellung Liescher Ringe,''
J.\ Fur die Reine and Angewandte Math.\ {\bf 177} 152 (1937).

\bibitem{May1972}
 J.~P.~May, 
 ``The geometry of iterated loop spaces,''
  Lecture Notes in Math. {\bf 271} (1972).

\bibitem{GinzburgKapranov1994}
V.~Ginzburg and M.~M.~Kapranov, 
``Koszul duality for operads,''
 Duke Math. J {\bf 76} 203 (1994) 
arXiv:0709.1228 [math.AG].

\bibitem{Kontsevich1994}
M.~Kontsevich, 
``Feynman diagrams and low-dimensional topology,''
 First European Congress of Mathematics II, Progr Math , vol 120, Birkh\"auser, Basel, 1994.

\bibitem{operads}
M.~Markl, S.~Shnider and J.~Stasheff,
``Operads in algebra, topolgy and physics,''
Mathematical surveys and monographs vol. 96, 2002.

\bibitem{Klyachko1974}
A.~A.~Klyachko,
``Lie elements in the tensor algebra,''
Sibirskii Matematicheskii Zhurnal (translation) {\bf 15} 1296 (1974).

 \bibitem{Brandt44}
 A.~Brandt, 
 ``The free Lie ring and Lie representations of the full linear group,''
  Trans.\ Amer.\ Math.\ Soc.\ {\bf 56} 528 (1944).

 \bibitem{Thrall42}
R.~M.~Thrall,  
``On symmetrized Kronecker powers and the structure of the free Lie ring,''
 Amer.\ J.\ Math.\ {\bf 64} 371 (1942).


\bibitem{Getzler94}
E.~Getzler,
 ``Operads and moduli spaces of genus 0 Riemann surfaces,''
 The moduli space of curves (Texel Island, 1994), 199-230,
Progr.\ Math.\ 129, Birkh\"auser Boston, Boston, MA (1995)
[alg-geom/9411004]. 

\bibitem{TyeZhang2010b} 
  H.~Tye and Y.~Zhang,
  ``Remarks on the identities of gluon tree amplitudes,''
  Phys.\ Rev.\ D {\bf 82}, 087702 (2010)
  arXiv:1007.0597 [hep-th].

\bibitem{BernDixonDunbarKosower94} 
  Z.~Bern, L.~J.~Dixon, D.~C.~Dunbar and D.~A.~Kosower,
  ``One loop n point gauge theory amplitudes, unitarity and collinear limits,''
  Nucl.\ Phys.\ B {\bf 425}, 217 (1994)
  [hep-ph/9403226]. 

\bibitem{GAP4}
  The GAP~Group, \emph{GAP -- Groups, Algorithms, and Programming, 
  Version 4.7.4}; 
  2014,
  \verb+(http://www.gap-system.org)+.
  
\end{thebibliography}

\end{document}